\documentclass[aps,amsmath,notitlepage,amssymb,prMaterial,letterpaper,longbibliography]{revtex4-1}
\usepackage{amsfonts}
\usepackage{stmaryrd}
\usepackage{bbm}
\usepackage{mathrsfs}
\usepackage{tipa}
\usepackage{amssymb}
\usepackage{txfonts}
\usepackage{graphicx}
\usepackage{dcolumn}
\usepackage{epstopdf}
\usepackage[colorlinks,linkcolor=blue,urlcolor=blue,citecolor=blue]{hyperref}
\usepackage{multirow}
\usepackage{subfigure}
\usepackage{url}

\begin{document}
\title{Layered semiconductor EuTe$_{4}$ with charge density wave order in square tellurium sheets}
\author{D.Wu$^{1}$}
\author{Q.M. Liu$^{1}$}
\author{S.L. Chen$^{1}$}
\author{G.Y. Zhong$^{2}$}
\author{J. Su$^{3}$}
\author{L.Y. Shi$^{1}$}
\author{L. Tong$^{1}$}
\author{G. Xu$^{2}$}
\email{gangxu@hust.edu.cn}
\author{P. Gao$^{1}$}
\email{P-gao@pku.edu.cn}
\author{N.L. Wang$^{1}$}
\email{nlwang@pku.edu.cn}
\affiliation{$^{1}$International Center for Quantum Materials, School of Physics, Peking University, Beijing 100871, China}
\affiliation{$^{2}$Wuhan National High Magnetic Field Center and School of Physics, Huazhong University of Science and Technology, Wuhan 430074, China}
\affiliation{$^{3}$College of Chemistry and Molecular Engineering,Peking University, Beijing 100871, China}

\begin{abstract}
We report a novel quasi-two dimensional compound of EuTe$_{4}$ hosting charge density waves (CDW) instability. The compound has a crystallographic structure in a orthorhombic space group \emph{Pmmn} (No.59) with cell parameters \emph{a} = 4.6347(2){\AA}, \emph{b} = 4.5119(2){\AA}, \emph{c} = 15.6747(10){\AA} at room temperature. The pristine structure contains consecutive near-square Te sheets separated by corrugated Eu-Te slabs. Upon cooling, the compound experiences a phase transition near 255 K. X-ray crystallographic analysis and transmission electron microscopy (TEM) measurements reveal strong structural distortions in the low temperature phase, showing a 1\emph{a} $\times$ 3\emph{b} $\times$ 2\emph{c} superstructure with a periodic formation of Te-trimers in the monolayer Te sheets, yielding evidence for the formation of CDW order. The charge transport properties show a semiconducting behavior in the CDW state. Density functional theory calculations reveals a Fermi surface nesting driven instability with a nesting vector in good agreement with the one observed experimentally. Our finding provides a promising system for the study of CDW driven 2D semiconducting mechanisms, which would shed a new light on exploring novel 2D semiconductors with collective electronic states.

\end{abstract}


\maketitle

\section{Introduction}
Charge density waves (CDWs) are collective electronic condensate arising from strong coupling of conduction electrons and the underlying lattice in low-dimensional metals \cite{RevModPhys.60.1129,doi:10.1080/00018732.2012.719674,doi:10.1143/PTP.121.1289}. The subject has generated considerable interest in condensed matter physics due to its important insight into electron-phonon interaction and its potential role in the phase diagram of superconducting cuprates  \cite{Wu2011,Kawasaki2017, Chang2012a}. Most CDW formations are driven by Fermi surface (FS) nesting effect, i.e., the matching of sections of FS to others by a wave vector \emph{\textbf{q}}, which is favorable to lower the electronic energy of a system. In quasi-1D CDW systems, the perfect nesting condition can be easily realized and the FS would be fully gapped, resulting in a semiconducting or insulating phase, as manifested in most quasi-1D materials. While for higher dimensional systems, the FS nesting could hardly remove all parts of the FSs. In this circumstance, the CDW energy gap opens up only at the nested parts of FSs. Due to the residual ungapped regions of FSs, the CDW phase tends to keep metallic. Though a few examples exhibit semiconducting 2D CDW states, their semiconducting property is not truly driven by the FS nesting but by Mott interaction \cite{PhysRevLett.94.036405,Sipos2008} or by other physical mechanisms \cite{PhysRevLett.88.226402,PhysRevLett.99.027404}.

The quasi-2D binary rare-earth pollytellurides \emph{R}Te$_{n}$ (where \emph{R} is a lanthanide, n=2, 2.5, 3) are well-known systems with FS nesting-driven CDW states \cite{doi:10.1021/ja7111405, PhysRevB.90.085105, PhysRevLett.81.886, PhysRevLett.93.126405,B201162J,Papoian2000, Malliakas2005,Schmitt1649,PhysRevLett.96.226401}. The common structure of these compounds contain consecutive square Te sheets separated by isolated corrugated \emph{R}-Te slabs. $R$ is normally trivalent in the compound, donating three electrons to the system. They completely fill the Te \emph{p} orbitals in the \emph{R}-Te slabs, but partially fill those Te \emph{p} orbitals in the planar Te sheets. In their CDW state, these compounds remain metallic because of the imperfect FS nesting. Among \emph{R}Te$_{n}$ series, LaTe$_{2}$ has been suggested to be a potential instance having semiconducting CDW state. However, the reported charge transport properties indicate that LaTe$_{2}$  is a bad-metal rather than an insulator \cite{MIN2002205,PhysRevB.72.085132,Kwon2000}. The detailed infrared spectroscopy studies on LaTe$_{2}$ are also not consistent with the entirely gapped FS in its CDW state\cite{Huang2012,PhysRevB.75.205133}.

Here we report a novel compound of quasi-2D divalent rare-earth telluride EuTe$_{4}$ which exhibits a striking semiconducting behavior in CDW state. The prime structure of this compound can be considered as derived from the LaTe$_{2}$-type structure by intercalating two more Te atomic sheets between the doublet Eu-Te corrugated slabs. Upon cooling, the x-ray crystallographic analysis and transmission electron microscopy (TEM) study reveal strong in-plane structural distortions, with a 1\emph{a} $\times$ 3\emph{b} superstructure modulation, yielding evidence for the formation of CDW order. The temperature-dependent charge transport measurements confirm the transition occurring around temperature T$_{c}$ = 255 K. After the transition, the state exhibits a narrow-gap semiconducting behavior, with the activation energy gap estimated to be $\sim$25 meV by the Arrhenius law. This semiconducting phase is further analyzed based on the Density Functional Theory (DFT) calculations. The modeling indicates that the FS topology favors a nesting vector along the \emph{b}-axis direction with a value of \emph{\textbf{q} }= \emph{\textbf{b}}$^{*}$/3 (where \emph{\textbf{b}}$^{*}$ = 2$\pi$/\emph{\textbf{b}}), which is quite well consistent with the experimental observations. Our result suggests a nesting driven CDW phase in EuTe$_{4}$, which lowers the electronic energy of the system and is responsible for the semiconducting properties.

\section{Results and discussion}

EuTe$_{4}$ single crystals were grown via the Te flux approach. High-purity Eu lumps (99.999\%) and Te granules (99.999\%) were mixed with a ratio of $\sim$ 1 : 15. The total weighted starting materials were sealed in an evacuated fused silica tube in high vacuum (10$^{-5}$ mbar) followed by heating at 850 $^{\circ}$C for 2 days in a muffle furnace. Then the furnace was slowly cooled to 415 $^{\circ}$C in 100 hours, and hold at this temperature for one weak then decanted using a centrifuge. The crystals are planar shaped with dark and mirror-like surfaces.

The crystallographic structure analysis of EuTe$_{4}$ single crystals gives a symmetry of \emph{Pmmn} (No.59) with cell parameters \emph{a} = 4.5119(2){\AA}, \emph{b} = 4.6347(2){\AA}, \emph{c} = 15.6747(10) {\AA} near room temperature (RT), as shown in Figure 1a. The structure can be considered as derived from the LaTe$_{2}$-type structure by intercalating two more Te atomic sheets into the doublet Eu-Te slabs (Figure S1). The quantitative energy dispersive X-Ray spectroscopy (EDX) analysis of the compounds (Figure S2) gives the atomic ratio of Eu:Te close to a stoichiometric 1:4, consistent with the crystal refinement results (Table 1). Figure 1b is an atomic-resolution high angle annular dark field (HAADF) scanning transmission electron microscope (STEM) image of EuTe$_{4}$, highlighting the structure of quintuple layer stacking. Between the adjacent quintuple layers, the nearest bond (Te-Te) is 3.37 {\AA}, which is much larger than the normal covalent Te-Te bond of 2.8 {\AA}, indicating weak inter-layer interactions.

A striking structural feature of EuTe$_{4}$ is the appearance of the regular near-square nets made of the monolayer Te atoms. Figure 1c depicts the four different crystallographic positions of Te atoms, verifying two inequivalent Te sheets of Te(3) and Te(4)-Te(5) networks. Within these Te sheets (Figure 1d), the Te-Te bonding is 3.2341(1) {\AA}- a typical hypervalent Te-Te bond length\cite{Papoian2000}, suggesting high propensity for structural distortions. In the title compound, the Eu atom is 9-coordinate in an square antiprismatic geometry, as shown in Figure 1e. The near-neighbouring Eu-Te bond length, having the value between 3.3125(6){\AA} (Eu-Te(2) ) and 3.517(2) {\AA}(Eu-Te(3)), are compatible with the global $R$-Te distances observed in $R$Te$_{n}$ families. Figure 1f shows the photograph of typical single crystals.

\begin{table}
  \caption{Crystal data and structure refinement for EuTe$_{4}$ at 289.98 K.}
  \label{tbl:290 K}
  \begin{tabular}{ll}
    \hline
    Empirical formula & EuTe$_{4}$  \\
    \hline
    Formula weight & 662.36   \\
    Temperature & 289.98(11) K  \\
    Crystal system & orthorhombic \\
    Space group & Pmmn \\
    unit cell dimens & \emph{a} = 4.5119(2){\AA}, $\alpha$ = 90$^{\circ}$ \\
      & \emph{b} = 4.6347(2){\AA}, $\beta$ = 90$^{\circ}$  \\
      & \emph{c} = 15.6747(10){\AA}, $\gamma$ = 90$^{\circ}$  \\
      Volume, Z  & 327.78(3) {\AA}$^{3}$, 2\\
      density(calcd) & 6.711 g/cm$^{3}$ \\
      absorp coeff & 26.872 mm$^{-1}$ \\
      F(000) & 542.0 \\
      Crystal size ( mm$^{3}$ ) & 0.15 $\times$ 0.1 $\times$ 0.02 \\
      Radiation & Mok$\alpha$ ($\lambda$ = 0.71073{\AA})\\
      2$\theta$ range for data collection   &   5.198$^{\circ}$  to 52.732$^{\circ}$  \\
      Index ranges & -5 $\leq$ h $\leq$ 5, -3 $\leq$ k $\leq$ 5, -19 $\leq$ l $\leq$ 19 \\
      Reflections collected & 2880 \\
      Independent reflections & 429 [ R$_{int}$ = 0.0481, R$_{sigma}$ = 0.0286 ] \\
      Data/restraints/parameters & 429/0/21 \\
      Goodness-of-fit on F$^{2}$ & 1.095 \\
      Final R indexes [I $>$=2$\sigma$ (I)] & R$_{1}$ = 0.0456, wR$_{2}$ = 0.1340  \\
      Final R indexes [all data] & R$_{1}$ = 0.0468, wR$_{2}$ = 0.1347  \\
      Largest diff. peak/hole / e ( {\AA}$^{3}$ ) & 4.28/-3.60 \\

    \hline
  \end{tabular}
\end{table}

\begin{table}
  \caption{Crystal data and structure refinement for EuTe$_{4}$ at 81 K.}
  \label{tbl:81 K}
  \begin{tabular}{ll}
    \hline
    Empirical formula & Eu$_{3}$Te$_{12}$  \\
    \hline
    Formula weight & 1987.08   \\
    Temperature & 81(2) K  \\
    Crystal system & orthorhombic \\
    Space group & P2$_{1cn}$ \\
    unit cell dimens & \emph{a} = 4.4898(3){\AA}, $\alpha$ = 90$^{\circ}$ \\
      & \emph{b} = 13.8903(11){\AA}, $\beta$ = 90$^{\circ}$  \\
      & \emph{c} = 31.268(2){\AA}, $\gamma$ = 90$^{\circ}$  \\
      Volume, Z  & 1950.0(2) {\AA}$^{3}$, 4\\
      density(calcd) & 6.768 g/cm$^{3}$ \\
      absorp coeff & 27.101 mm$^{-1}$ \\
      F(000) & 3252.0 \\
      Crystal size ( mm$^{3}$ ) & 0.272 $\times$ 0.097 $\times$ 0.094 \\
      Radiation & Mok$\alpha$ ($\lambda$ = 0.71073{\AA})\\
      2$\theta$ range for data collection   &   3.922$^{\circ}$  to 52.738$^{\circ}$  \\
      Index ranges & -5 $\leq$ h $\leq$ 5, -17 $\leq$ k $\leq$ 17, -39 $\leq$ l $\leq$ 39 \\
      Reflections collected & 18379 \\
      Independent reflections & 3908 [ R$_{int}$ = 0.0660, R$_{sigma}$ = 0.0530 ] \\
      Data/restraints/parameters & 3908/85/131 \\
      Goodness-of-fit on F$^{2}$ & 1.073 \\
      Final R indexes [I $>$=2$\sigma$ (I)] & R$_{1}$ = 0.0707, wR$_{2}$ = 0.1953 \\
      Final R indexes [all data] & R$_{1}$ = 0.0827, wR$_{2}$ = 0.2032  \\
      Largest diff. peak/hole / e ( {\AA}$^{3}$ ) & 4.38/-4.41 \\

    \hline
  \end{tabular}
\end{table}

\begin{figure}

  \includegraphics[width=10cm]{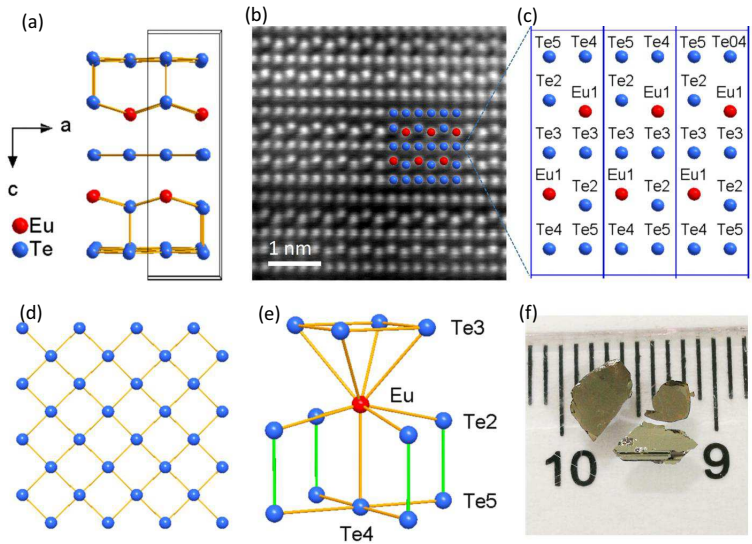}\\
  \caption{(a)The crystallographic structure of EuTe$_{4}$ along the [010] direction with alternating Eu-Te slabs and square Te atomic sheets available at RT. (b) The HAADF-STEM image along [010] direction matches perfectly to the atomic positions refined from the X-ray diffraction data. (c) Depicts the crystallographic positions of Te and Eu atoms. (d) The square Te atomic networks along the [001] direcition. (e) The 9-coordinated Eu in a square antiprismatic geometry. The covalent bond of Te(2)-Te(5) is outlined in green. (f) Images of the as-synthesized single crystals showing shinning surfaces, the scale is in mm.}
\end{figure}

Hinted by the binary rare-earth polytellurides \emph{R}Te$_{n}$ (n=2, 2.5, 3), the presence of square Te sheets in the structure suggests that the compound is susceptible to Peierls instability, as the distorted structure is more energetically stable than the ideal square net structure\cite{doi:10.1021/ja00235a021,Papoian2000}. To reveal further characteristics of the structural distortions, the low temperature single-crystal X-ray diffraction analysis was employed. The result indicates that the system evolves into a modulated structural phase at cooling temperatures, showing a new structural symmetry \emph{Pna21} (Table 2). An in-plane commensurate supercell 1\emph{a} $\times$ 3\emph{b} is constructed. Figure 2a and 2b depict such the fragments of the modulated structures of the Te sheets projected onto the \emph{ab} plane at 81K. The modulated pattern can be seen as a sequence of V-shaped trimers in the planar Te nets. This converts a situation of all weak bonding in the undistorted monolayer Te sheets to a situation of some strong and some weak bondings in the distorted Te sheets. For instance, Te(3) monolayer sheets (Figure 2a), the minimum, maximum and average Te-Te distance after the distortion are 3.0064(4) {\AA}, 3.4701(4) {\AA} and 3.1824(7) {\AA}, respectively. The similar distortions also exist in Te(4)-Te(5) networks (Figure 2b),with the minimum, maximum and average Te-Te distance of 2.8364(4) {\AA}, 3.4919(5) {\AA} and 3.2650(1) {\AA}, respectively. Corresponding to the displacement distortions in the Te sheets, the in plane Te-Te bond angle are strongly deformed as well. For those Te-trimers, the bond angle is 99.89$^{\circ}$ in Te(3) sheets and 100.585$^{\circ}$ in Te(4)-Te(5) sheets, showing strong deviations from the right angle. In the distorted structure, the Te-trimers in Te(4)-Te(5) sheets combine with Te(2) atoms through the covalent Te-bonds give rise to the 3D structural formations of Te tetramers and pentamers, as displayed in Fig. 2c, 2d, and 2e.

The low-temperature TEM study was employed to reveal further the morphologic distortions in EuTe$_{4}$ system.
Figure 3a shows the selected area electron diffraction (SAED) pattern of EuTe$_{4}$ at 95 K along the [001] zone axis. The superlattice spots can be clearly differentiated only along [010] direction, showing a \emph{\textbf{q}}-vector $\sim$ 0.33 \emph{\textbf{b}}$^{*}$. The modulated lattice remains in an orthorhombic symmetry, consistent with the X-ray diffraction data. The TEM study also confirms that there is no superstructure existing for the pristine lattice at room temperature (Figure 3b).

The valence state of Eu in EuTe$_{4}$ was analyzed by the magnetic property studies. Figure 3c shows the magnetic susceptibility of EuTe$_{4}$, the data recognize an antiferromagnetic phase transition (T$_{N}$ = 7.1 K), arising from the in-plane correlation of magnetic moments of Eu ions. Above the Neel temperature, the  susceptibility can be modeled using the Curie - Weiss law [$\chi$ = $\chi$$_{0}$ + C/(T - $\Theta$)] (Figure S3). The derived effective magnetic moment of 7.65 $\mu_{B}$ per Eu agrees well with the theoretical value of 7.9 $\mu_{B}$ for free Eu$^{2+}$ ions. As will be discussed in the following DFT paragraph, the 4f electrons of Eu are strongly localized around -1.5 eV below the FS and the electronic bands near the FS are mainly from Te 5p orbitals. Therefore, the localized 4f electrons of Eu$^{2+}$ are considered having no influence on the CDW order. Furthermore, owing to large magnetic moment, Eu$^{2+}$ dominates the magnetic susceptibility in EuTe$_{4}$ material. In our measurements, there is no anomaly observed in the measured magnetic susceptibility near the CDW transition, suggesting that the CDW's contribution of susceptibility is too small to be detected.

A well-defined semiconductive behavior was found for EuTe$_{4}$. As is recognized from the temperature dependent in-plane electrical resistivity, for both cooling and heating process (Figure 3d). The large temperature hysteresis yields evidence for a first-order phase transition. The transition temperature T$_{c}$, defined from the resistivity kink of cooling process, is $\sim$ 255 K. Above T$_{c}$ and as cooling, the resistivity shows a positive temperature-dependence and has a value $\sim$ 1.1 $\times$ 10$^{-2}$ $\Omega$ cm at RT, indicating a bad-metal characteristic. While below T$_{c}$, the resistivity manifests a thermally activated semiconductive behavior. The slope of the logarithm resistivity vs 1000/T plot at low temperatures gives an activation energy of $\sim$ 25 meV (see Figure S4). The resistivity also shows a kink at 50K, which might be related to a further structural change in the system beyond our experimental determination. Differential scanning calorimetry on a large number of crystals shows a kink around the same temperature, consistent with the existence of the first-order phase transition in the bulk (Figure 3e).

\begin{figure}
  \centering
  \includegraphics[width=10cm]{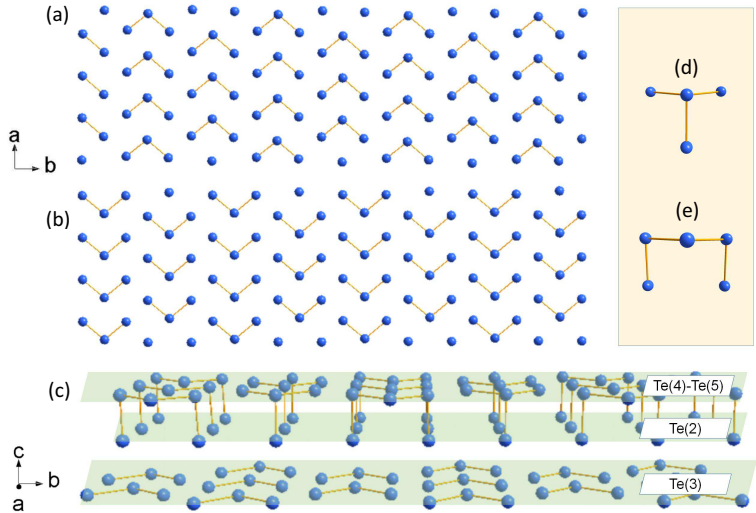}\\
  \caption{Distorted tellurium sheets of EuTe$_{4}$ at cooling temperature of 81 K. (a) The distortion in Te(4)-Te(5) sheet, a bond length threshold is set as 2.988 {\AA}. The sequence of V-shaped Te-trimers are clearly recognized. (b) Distortions in Te(3) net for a bond threshold 3.015 {\AA}, showing the similar formation of Te-trimers . (c) The in-plane Te-trimers in Te(4)-Te(5) sheet combining with Te(2) sheet through strong Te-Te covalent bond, forming into 3D Te-oligomers of Te tetramers (d) and pentamers (e).}
\end{figure}

Hall effect measurements were conducted on the same single crystal from which the resistivity was measured (Figure 3f), which give rise to a p-type charge conduction with carrier concentrations estimated to be 3.1 $\times$ 10$^{19}$ cm$^{-3}$, 1.9 $\times$ 10$^{19}$ cm$^{-3}$, 6.2 $\times$ 10$^{18}$ cm$^{-3}$ and 3.7 $\times$ 10$^{18}$ cm$^{-3}$ at 350 K, 300 K, 240 K, 220 K respectively based on one-type carrier model. The calculated carriers mobility, for instance, is $\sim$ 36 cm$^{2}$ V$^{-1}$ s$^{-1}$  at 300K and $\sim$ 42 cm$^{2}$ V$^{-1}$ s$^{-1}$ at 220 K. These results imply that EuTe$_{4}$ is a semiconductor having low carrier concentrations, in contrast to the common trivalent rare-earth species $R$Te$_{n}$ which are metals with higher carriers densities \cite{MIN2002205,PhysRevB.73.033101}.

In order to understand the origin of the structural phase transition, we have performed the electronic structures calculations based on the RT structure of EuTe$_{4}$ by the density functional theory (DFT) method. The BSTATE (Beijing Simulation Tool of Atomic TEchnology) package\cite{fang2002structural} with plane-wave pseudopotential method implemented was used for DFT calculations, all results are double-checked by the projector augmented wave method\cite{PhysRevB.50.17953,PhysRevB.59.1758} implemented in VASP package\cite{KRESSE199615}. The calculated total and projected density of states (DOS and PDOS) for the nonmagnetic state are plotted in Figure 4a, which shows that the states between $-0.5$ eV and $0.5$ eV are mostly contributed by the $5p_{x}$ and $5p_{y}$ orbitals of Te atoms. In order to figure out which kinds of Te atoms dominate near the Fermi level, the projected band structures for different kinds of Te atoms are plotted in Figure 4b, which indicate that the low-energy bands can be schematically separated into two parts. The narrow bands below $-1$ eV are mostly from $5p_{x}$ and $5p_{y}$ orbitals of Te(2) atoms, while the dispersive bands near the Fermi level are mainly contributed by the $5p_{x}$ and $5p_{y}$ orbitals from Te(3), Te(4) and Te(5) atoms. Such results demonstrate a metallic and anisotropic electronic structures for the RT EuTe$_{4}$, where the metallic electrons mainly come from the $5p_{x}$ and $5p_{y}$ orbitals of the Te square sheets, \emph{i.e.}, Te(3), Te(4) and Te(5) atoms.

\begin{figure}
  \centering
  \includegraphics[width=15cm]{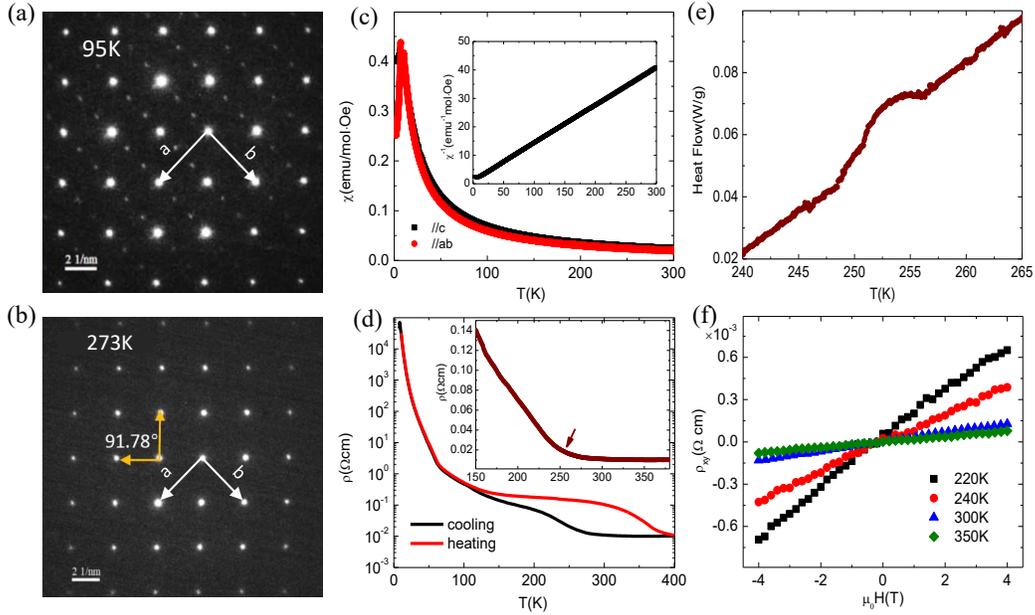}\\
  \caption{Magnetic and charge transport properties of EuTe$_{4}$. (a)The SAED pattern viewing along the [00l] zone axis from thin microcrystals of EuTe$_{4}$ at 95 K. Clear superlattice spots along only in b-axis direction with a modulation \textbf{q}-vector $\sim$ 0.33 \textbf{b}$^{*}$ can be observed. (b) The SAED pattern viewing along the [00l] zone axis from thin microcrystals of EuTe$_{4}$ at 273 K. (c) The temperature dependence of the magnetic susceptibility recognize a para-magnetic behavior above the N¨¦el temperature ($\sim$ 7.1 K). (d) The temperature dependence of the electrical resistivity see a first-order phase transition showing a big temperature hysteresis.
  The inset shows for the cooling process, the resistivity kink around $\sim$ 255 K. (e) Differential scanning calorimetry on a large number of crystals shows a kink locating around 255 K, confirming the existence of the phase transition in the bulk. (f) In-plane Hall resistivity. The positive slop indicates hole-dominated carriers.}
\end{figure}

Such anisotropy can also be reflected by the calculated two-dimensional like Fermi surfaces (FSs), as shown in Figure 4c. There are three types of FSs in Figure 4c, three concave-square hole FSs around the $\Gamma$ point (blue FSs surrounding $\Gamma$), six spindle-shape electron FSs around the X and Y points (three red FSs surrounding X and three red FSs surrounding Y), and three convex-square hole FSs around the S point (blue FSs surrounding the corner). It's obvious that there are two vectors that can induce large FSs nesting effect, \emph{i.e.}, by shifting the FSs of $\textbf{q}=\textbf{b}_{1}/3$ and $\textbf{q}=\textbf{b}_{2}/3$, both can lead to the significant overlap of the FSs. In order to clarify which vector is favored, we calculated the Lindhard response function\cite{dong2008competing,PhysRevB.77.165135}, and plotted the normalized two-dimensional Lindhard response function $\chi_{0}(\textbf{q})$ in Figure 4d, where the value of $\chi_{0}$, \emph{i.e.}, the brightness in Figure 4d, can be used to quantitatively estimate the strength of the nesting effect. The calculated $\chi_{0}(\textbf{q})$ is more strongly peaked at $\textbf{q}=\textbf{b}_{2}/3$ than that at $\textbf{q}=\textbf{b}_{1}/3$, which can also be demonstrated by the normalized one-dimensional $\chi_{0}(\textbf{q})$ as shown in Figure 4e. In Figure 4e, we have also plotted the normalized one-dimensional $\chi_{0}(\textbf{q})$ along different vectors, such as $\textbf{Q}=\textbf{b}_{3}$, $\textbf{Q}=\textbf{b}_{1}+\textbf{b}_{2}$ and $\textbf{Q}=\textbf{b}_{2}+1.5\textbf{b}_{3}$ (whose $1/3$ corresponding to 1$\times$3$\times$2 reconstruction in the real space). We find that the biggest $\chi_{0}(\textbf{q})$ appears at $\textbf{q}=\textbf{b}_2/3$ or $\textbf{q}=(\textbf{b}_{2}+1.5\textbf{b}_{3})/3$, which demonstrates two important facts. Firstly, the FSs of RT EuTe$_{4}$ are quite two-dimensional. Secondly, the nesting effect induced by the $\textbf{b}_{2}/3$ shifting is responsible for the 1$\times$3 reconstruction of the Te square sheets, which agrees well with the experimental observation of the in-plane supercell vector.

\begin{figure}
  \centering
  \includegraphics[width=14cm]{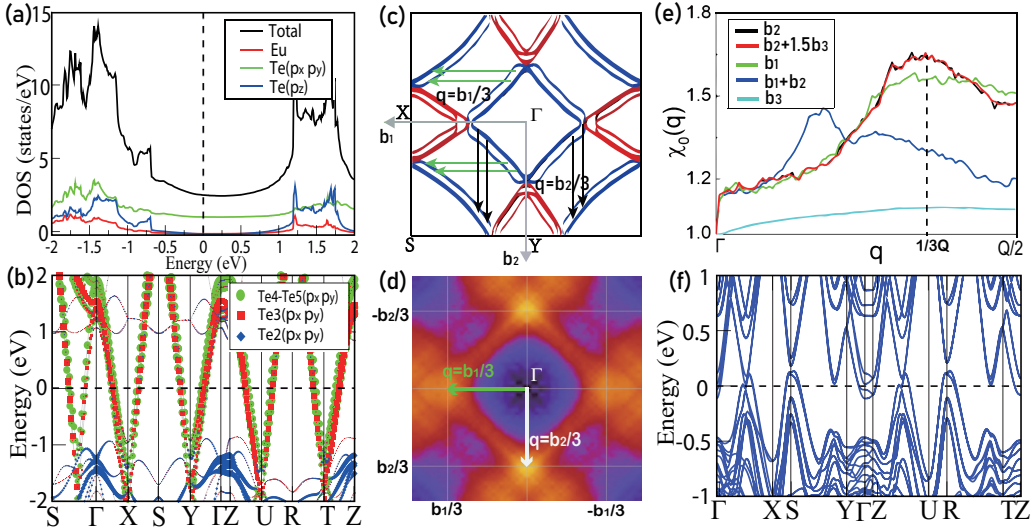}\\
  \caption{DFT calculations for EuTe$_{4}$ (a) Calculated DOS and PDOS for the RT structure. (b) The projected band structures for different kinds of Te atoms. The size of the green circles, red squares and blue diamonds represent the weight of $p_{x}$ and $p_{y}$ orbitals for Te(4)-Te(5), Te(3) and Te(2) atoms, respectively. (c) Top view of the calculated FSs for the RT structure of EuTe$_{4}$, where two important nesting vectors, $\textbf{q}=\textbf{b}_{2}/3$ or $\textbf{q}=\textbf{b}_{1}/3$ are indicated by green and black arrows, respectively. The vectors $\textbf{b}_{1}= 2\pi/\textbf{a}$, $\textbf{b}_{2}= 2\pi/\textbf{b}$ and $\textbf{b}_{3}= 2\pi/c$. (d) and (e) The normalized two-dimensional Lindhard response function $\chi_{0}(\textbf{q})$ in $\textbf{b}_{1}$$\textbf{b}_{2}$-plane, and the one-dimensional normalized $\chi_{0}(\textbf{q})$ along different vectors. (f) Band structures calculated based on the experimental low-temperature structure and the GGA type of the exchange-correlation potential.}
\end{figure}

The good two-dimensional character of the FSs makes $\textbf{b}_{2}/3$ a nearly perfect nesting vector. In Figure S5b, we plot a schematic of FSs nesting by $\textbf{q}=\textbf{b}_{2}/3$ vector, where all the original FSs (solid curves in red and blue) in the reconstructed Brillouin zone (BZ, the area between the black dashes) are nested with the folded FSs (dashed curves in red and blue) almost perfectly. Such strong nesting effect by the vector $\textbf{q}=\textbf{b}_{2}/3$ suggests that most the FSs will be gapped after the ($1\times3$) superlattice reconstruction, which is also confirmed by our electric structures calculations on the low-temperature structure (1$\times$3$\times$2). In Figure 4f, we plot the band structures calculated based on the experimental low-temperature structure and the generalized gradient approximation (GGA)\cite{PhysRevLett.77.3865} type of the exchange-correlation potential. Comparing with Figure 4a-4c, the FSs is much smaller for the low-temperature structure, and its DOS at the Fermi level is reduced to 0.63 states/(eV f.u.) (1.38 states/(eV f.u.) for the RT structure), indicating that most FSs are gapped by the (1$\times$3) reconstruction. And the total energy of LT structure is about 4 meV/f.u. lower than that of the RT structure due to the FSs nesting. As we know, GGA usually overestimates the overlap of the conduction bands and the valence bands. The nonlocal Heyd-Scuseria-Ernzerhof (HSE06) hybrid functional calculations \cite{Heyd2003} can partially overcome this flaw. We have performed HSE06 calculations on the low-temperature structure of EuTe$_{4}$. The DOS at the Fermi level, as shown in Figure S5c, is further reduced to 0.35 states/(eV f.u.). This may explain why the transport measurement of EuTe$_{4}$ exhibits an insulating behavior at low temperature.

\section{conclusion}

A new compound of binary rare-earth telluride EuTe$_{4}$ is synthesized, which contains Te atomic quasi-square sheets. Eu is divalent, an only exceptional species among the layered \emph{R}Te$_{n}$ (n $\geq$ 2) polytellurides where \emph{R} is normally trivalent, to the best of our knowledge. At RT, EuTe$_{4}$ has a crystal structure containing two crystallographic inequivalent Te quasi-square sheets, each of them has strong CDW modulations after the Peierls transition. Unlike common 2D CDW systems which have metallic CDW state, EuTe$_{4}$ has a semiconducting CDW phase. Detailed DFT calculations indicate that the FS topology of EuTe$_{4}$ favors a nesting vector with a value of \emph{\textbf{q} }= \emph{\textbf{b}}$^{*}$/3 , corresponding closely with the experimental detected superstructure vector. Our results suggest a nesting driven CDW phase for EuTe$_{4}$ material at cooling temperatures, which is responsible for the semiconducting state.

\section{acknowledgement}

This work was supported by the National Science Foundation
of China (Nos. 11888101, 51502007 and 51672007), the National Key Research and Development Program of China (nos. 2017YFA0302904, 2016YA0300902, 2016YFA0300903, 2016YFA0300804 and 2018YFA0307000). We gratefully acknowledge Microscopy Laboratory in Peking University for the use of Cs corrected electron microscope and in situ TEM platform.

\bibliography{EuTe4}

\end{document}